# SPARSE ESTIMATION OF LARGE COVARIANCE MATRICES VIA A NESTED LASSO PENALTY

By Elizaveta Levina,[1] Adam Rothman and Ji Zhu[2]

*University of Michigan*

The paper proposes a new covariance estimator for large covariance matrices when the variables have a natural ordering. Using the Cholesky decomposition of the inverse, we impose a banded structure on the Cholesky factor, and select the bandwidth adaptively for each row of the Cholesky factor, using a novel penalty we call nested Lasso. This structure has more flexibility than regular banding, but, unlike regular Lasso applied to the entries of the Cholesky factor, results in a sparse estimator for the inverse of the covariance matrix. An iterative algorithm for solving the optimization problem is developed. The estimator is compared to a number of other covariance estimators and is shown to do best, both in simulations and on a real data example. Simulations show that the margin by which the estimator outperforms its competitors tends to increase with dimension.

**1. Introduction.** Estimating covariance matrices has always been an important part of multivariate analysis, and estimating large covariance matrices (where the dimension of the data $p$ is comparable to or larger than the sample size $n$) has gained particular attention recently, since high-dimensional data are so common in modern applications (gene arrays, fMRI, spectroscopic imaging, and many others). There are many statistical methods that require an estimate of a covariance matrix. They include principal component analysis (PCA), linear and quadratic discriminant analysis (LDA and QDA) for classification, regression for multivariate normal data, inference about functions of the means of the components (e.g., about the mean response curve in longitudinal studies), and analysis of independence and conditional independence relationships between components in graphical models. Note that in many of these applications (LDA, regression, conditional independence analysis) it is not the population covariance $\Sigma$ itself

Received March 2007; revised September 2007.

[1]Supported in part by NSF Grant DMS-05-05424 and NSA Grant MSPF-04Y-120.

[2]Supported in part by NSF Grants DMS-05-05432 and DMS-07-05532.

that needs estimating, but its inverse $\Sigma^{-1}$, also known as the precision or concentration matrix. When $p$ is small, an estimate of one of these matrices can easily be inverted to obtain an estimate of the other one; but when $p$ is large, inversion is problematic, and it may make more sense to estimate the needed matrix directly.

It has long been known that the sample covariance matrix is an extremely noisy estimator of the population covariance matrix when $p$ is large, although it is always unbiased [Dempster (1972)]. There is a fair amount of theoretical work on eigenvalues of sample covariance matrices of Gaussian data [see Johnstone (2001) for a review] that shows that unless $p/n \to 0$, the eigenvalues of the sample covariance matrix are more spread out than the population eigenvalues, even asymptotically. Consequently, many alternative estimators of the covariance have been proposed.

Regularizing large covariance matrices by Steinian shrinkage has been proposed early on, and is achieved by either shrinking the eigenvalues of the sample covariance matrix [Haff (1980); Dey and Srinivasan (1985)] or replacing the sample covariance with its linear combination with the identity matrix [Ledoit and Wolf (2003)]. A linear combination of the sample covariance and the identity matrix has also been used in some applications— for example, as original motivation for ridge regression [Hoerl and Kennard (1970)] and in regularized discriminant analysis [Friedman (1989)]. These approaches do not affect the eigenvectors of the covariance, only the eigenvalues, and it has been shown that the sample eigenvectors are also not consistent when $p$ is large [Johnstone and Lu (2007)]. Hence, shrinkage estimators may not do well for PCA. In the context of a factor analysis model, Fan et al. (2008) developed high-dimensional estimators for both the covariance and its inverse.

Another general approach is to regularize the sample covariance or its inverse by making it sparse, usually by setting some of the off-diagonal elements to 0. A number of methods exist that are particularly useful when components have a natural ordering, for example, for longitudinal data, where the need for imposing a structure on the covariance has long been recognized [see Diggle and Verbyla (1998) for a review of the longitudinal data literature]. Such structure often implies that variables far apart in this ordering are only weakly correlated. Banding or tapering the covariance matrix in this context has been proposed by Bickel and Levina (2004) and Furrer and Bengtsson (2007). Bickel and Levina (2007) showed consistency of banded estimators under mild conditions as long as $(\log p)/n \to 0$, for both banding the covariance matrix and the Cholesky factor of the inverse discussed below. They also proposed a cross-validation approach for selecting the bandwidth.

Sparsity in the inverse is particularly useful in graphical models, since zeroes in the inverse imply a graph structure. Banerjee et al. (2006) and Yuan and Lin (2007), using different semi-definite programming algorithms, both achieve sparsity by penalizing the normal likelihood with an $L_1$ penalty imposed directly on the elements of the inverse. This approach is computationally very intensive and does not scale well with dimension, but it is invariant under variable permutations.

When a natural ordering of the variables is available, sparsity in the inverse is usually introduced via the modified Cholesky decomposition [Pourahmadi (1999)],

$$\Sigma^{-1} = T^{\top} D^{-1} T.$$

Here $T$ is a lower triangular matrix with ones on the diagonal, $D$ is a diagonal matrix, and the elements below diagonal in the $i$th row of $T$ can be interpreted as regression coefficients of the $i$th component on its predecessors; the elements of $D$ give the corresponding prediction variances.

Several approaches to introducing zeros in the Cholesky factor $T$ have been proposed. While they are not invariant to permutations of variables and are thus most natural when variables are ordered, they do introduce shrinkage, and in some cases, sparsity, into the estimator. Wu and Pourahmadi (2003) propose a $k$-diagonal (banded) estimator, which is obtained by smoothing along the first $k$ sub-diagonals of $T$, and setting the rest to 0. The number $k$ is chosen via an AIC penalty on the normal likelihood of the data. The resulting estimate of the inverse is also $k$-banded. Wu and Pourahmadi (2003) showed element-wise consistency of their estimator (although that is a property shared by the sample covariance matrix), and Bickel and Levina (2007) showed that banding the Cholesky factor produces a consistent estimator in the matrix $L_2$ norm under weak conditions on the covariance matrix, the most general theoretical result on banding available to date. Huang et al. (2006) proposed adding an $L_1$ penalty on the elements of $T$ to the normal likelihood, which leads to Lasso-type shrinkage of the coefficients in $T$, and introduces zeros in $T$ which can be placed in arbitrary locations. This approach is more flexible than banding, but the resulting estimate of the inverse may not have any zeros at all, hence, the sparsity is lost. No consistency results are available for this method. A related Bayesian approach [Smith and Kohn (2002)] introduces zeros in the Cholesky factor via a hierarchical prior, while Wong et al. (2003) use a prior that allows elements of the inverse itself to be zero.

Our approach, which we will call adaptive banding in contrast to regular banding, also relies on the Cholesky decomposition and a natural ordering of the variables. By introducing a novel nested Lasso penalty on the coefficients of regressions that form the matrix $T$, we select the best model that regresses the $j$th variable on its $k$ closest predecessors, but, unlike in simple banding, we allow $k = k_j$ to depend on $j$. The resulting structure of the Cholesky factor is illustrated in Figure 1(c). It is obviously more flexible than banding, and hence, should produce a better estimate of the covariance by being better able to adapt to the data. Unlike the Lasso of Huang et al. (2006), adaptive banding preserves sparsity in the resulting estimate of the inverse, since the matrix $T$ is still banded, with the overall $k = \max_j k_j$. We show that adaptive banding, in addition to preserving sparsity, outperforms the estimator of Huang et al. (2006) in simulations and on real data. One may also reasonably expect that as long as the penalty tuning parameter is selected appropriately, the theoretical consistency results established for banding in Bickel and Levina (2007) will hold for adaptive banding as well.

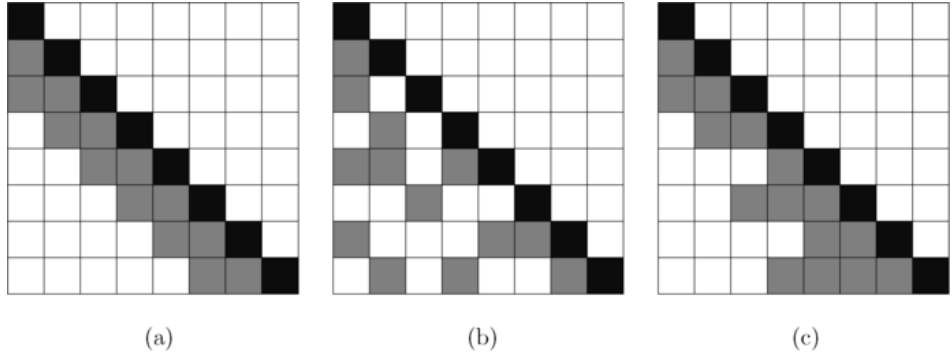

Fig. 1. *The placement of zeros in the Cholesky factor $T$:* (a) *Banding;* (b) *Lasso penalty of Huang et al.;* (c) *Adaptive banding.*

The rest of the paper is organized as follows: Section 2 summarizes the penalized estimation approach in general, and presents the nested Lasso penalty and the adaptive banding algorithm, with a detailed discussion of optimization issues. Section 3 presents numerical results for adaptive banding and a number of other methods, for simulated data and a real example. Section 4 concludes with discussion.

**2. Methods for penalized estimation of the Cholesky factor.** For the sake of completeness, we start from a brief summary of the formal derivation of the Cholesky decomposition of $\Sigma^{-1}$. Suppose we have a random vector $\mathbf{X} = (X_1, \ldots, X_p)^\top$, with mean $\mathbf{0}$ and covariance $\Sigma$. Let $X_1 = \varepsilon_1$ and, for $j > 1$, let

$$X_j = \sum_{l=1}^{j-1} \phi_{jl} X_l + \varepsilon_j,$$

where $\phi_{jl}$ are the coefficients of the best linear predictor of $X_j$ from $X_1, \ldots, X_{j-1}$ and $\sigma_j^2 = \mathrm{Var}(\varepsilon_j)$ the corresponding residual variance. Let $\Phi$ be the lower triangular matrix with $j$th row containing the coefficients $\phi_{jl}$, $l = 1, \ldots, j-1$, of the $j$th regression (1). Note that $\Phi$ has zeros on the diagonal. Let $\varepsilon = (\varepsilon_1, \ldots, \varepsilon_p)^\top$, and let $D = \mathrm{diag}(\sigma_j^2)$ be a diagonal matrix with $\sigma_j^2$ on the diagonal. Rewriting (1) in matrix form gives

$$\varepsilon = (I - \Phi)\mathbf{X}, \tag{2}$$

where $I$ is the identity matrix. It follows from standard regression theory that the residuals are uncorrelated, so taking covariance of both sides of (2) gives

$$D = (I - \Phi)\Sigma(I - \Phi)^\top.$$

Letting $T = I - \Phi$, we can now write down the modified Cholesky decompositions of $\Sigma$ and $\Sigma^{-1}$:

$$\Sigma = T^{-1}D(T^{-1})^\top, \qquad \Sigma^{-1} = T^\top D^{-1} T. \tag{3}$$

Note that the only assumption on $\mathbf{X}$ was mean $\mathbf{0}$; normality is not required to derive the Cholesky decomposition.

The natural question is how to estimate the matrices $T$ and $D$ from data. The standard regression estimates can be computed as long as $p \leq n$, but in high-dimensional situations one expects to do better by regularizing the coefficients in $T$ in some way, for the same reasons one achieves better prediction from regularized regression [Hastie et al. (2001)]. If $p > n$, the regression problem becomes singular, and some regularization is necessary for the estimator to be well defined.

2.1. *Adaptive banding with a nested Lasso penalty.* The methods proposed by Huang et al. (2006) and Wu and Pourahmadi (2003) both assume the data $\mathbf{x}_i$, $i = 1, \ldots, n$, are sampled from a normal distribution $\mathcal{N}(\mathbf{0}, \Sigma)$ and use the normal likelihood as the loss function. As the derivation above shows, the normality assumption is not necessary for estimating covariance using the Cholesky decomposition. We start, however, with the normal likelihood as the loss function and demonstrate how a new penalty can be applied to produce an adaptively banded estimator.

The negative log-likelihood of the data, up to a constant, is given by

$$\ell(\Sigma, \mathbf{x}_1, \ldots, \mathbf{x}_n) = n \log |\Sigma| + \sum_{i=1}^{n} \mathbf{x}_i^\top \Sigma^{-1} \mathbf{x}_i$$

$$= n \log |D| + \sum_{i=1}^{n} \mathbf{x}_i^\top T^\top D^{-1} T \mathbf{x}_i. \tag{4}$$

The negative log-likelihood can be decomposed into

$$\ell(\Sigma, \mathbf{x}_1, \ldots, \mathbf{x}_n) = \sum_{j=1}^{p} \ell_j(\sigma_j, \phi_{\mathbf{j}}, \mathbf{x}_1, \ldots, \mathbf{x}_n),$$

where

$$(5) \qquad \ell_j(\sigma_j, \phi_{\mathbf{j}}, \mathbf{x}_1, \ldots, \mathbf{x}_n) = n \log \sigma_j^2 + \sum_{i=1}^{n} \frac{1}{\sigma_j^2} \left( x_{ij} - \sum_{l=1}^{j-1} \phi_{jl} x_{il} \right)^2.$$

Minimizing (4) is equivalent to minimizing each of the functions $\ell_j$ in (5), which is in turn equivalent to computing the best least squares fit for each of the regressions (1).

Wu and Pourahmadi (2003) suggested using an AIC or BIC penalty to select a common order for the regressions (1). They also subsequently smooth the sub-diagonals of $T$, and their method's performance depends on the exact choice of the smoother and the selection of the smoothing parameters as much as on the choice of order. This makes a direct comparison to Huang et al. (2006) and our own method difficult. Bickel and Levina (2007) proposed a cross-validation method for selecting the common order for the regressions, and we will use their method for all the (nonadaptive) banding results below.

Huang et al. (2006) proposed adding a penalty to (4) and minimizing

$$(6) \qquad \ell(\Sigma, \mathbf{x}_1, \ldots, \mathbf{x}_n) + \lambda \sum_{j=2}^{p} P(\phi_j),$$

where the penalty $P$ on the entries of $\phi_j = (\phi_{j1}, \ldots, \phi_{j,j-1})$ is

$$(7) \qquad P(\phi_j) = \|\phi_j\|_d^d,$$

and $\| \cdot \|_d$ is the $L_d$ vector norm with $d = 1$ or $2$. The $L_2$ penalty ($d = 2$) does not result in a sparse estimate of the covariance, so we will not focus on it here. The $L_1$ penalty ($d = 1$), that is, the Lasso penalty [Tibshirani (1996)], results in zeros irregularly placed in $T$ as shown in Figure 1(b), which also does not produce a sparse estimate of $\Sigma^{-1}$. Again, minimizing (6) is equivalent to separately minimizing

$$(8) \qquad \ell_j(\sigma_j, \phi_{\mathbf{j}}, \mathbf{x}_1, \ldots, \mathbf{x}_n) + \lambda P(\phi_j),$$

with $P(\phi_1) = 0$.

We propose replacing the $L_1$ penalty $\lambda \sum_{l=1}^{j-1} |\phi_{jl}|$ with a new *nested* Lasso penalty,

$$(9) \qquad J_0(\phi_j) = \lambda \left( |\phi_{j,j-1}| + \frac{|\phi_{j,j-2}|}{|\phi_{j,j-1}|} + \frac{|\phi_{j,j-3}|}{|\phi_{j,j-2}|} + \cdots + \frac{|\phi_{j,1}|}{|\phi_{j,2}|} \right),$$

where we define $0/0 = 0$. The effect of this penalty is that if the $l$th variable is not included in the $j$th regression ($\phi_{jl} = 0$), then all the subsequent variables ($l - 1$ through 1) are also excluded, since giving them nonzero coefficients would result in an infinite penalty. Hence, the $j$th regression only uses $k_j \leq$ $j - 1$ closest predecessors of the $j$th coordinate, and each regression has a different order $k_j$.

The scaling of coefficients in (9) could be an issue: the sole coefficient $\phi_{j,j-1}$ and the ratios $\frac{|\phi_{j,t}|}{|\phi_{j,t+1}|}$ can, in principle, be on different scales, and penalizing them with the same tuning parameter $\lambda$ may not be appropriate. In situations where the data have natural ordering, the variables are often measurements of the same quantity over time (or over some other index, e.g., spatial location or spectral wavelength), so both the individual coefficients $\phi_{j,t}$ and their ratios are on the order of 1; if variables are rescaled, in this context they would all be rescaled at the same time (e.g., converting between different units).

However, the nested Lasso penalty is of independent interest and may be used in other contexts, for example, for group variable selection. To address the scaling issue in general, we propose two easy modifications of the penalty (9):

$$(10) \qquad J_1(\phi_j) = \lambda \left( \frac{|\phi_{j,j-1}|}{|\hat{\phi}^*_{j,j-1}|} + \frac{|\phi_{j,j-2}|}{|\phi_{j,j-1}|} + \frac{|\phi_{j,j-3}|}{|\phi_{j,j-2}|} + \cdots + \frac{|\phi_{j,1}|}{|\phi_{j,2}|} \right),$$

$$(11) \qquad J_2(\phi_j) = \lambda_1 \sum_{t=1}^{j-1} |\phi_{j,t}| + \lambda_2 \sum_{t=1}^{j-2} \frac{|\phi_{j,t}|}{|\phi_{j,t+1}|},$$

where $\hat{\phi}^*_{j,j-1}$ is the coefficient from regressing $X_j$ on $X_{j-1}$ *alone*. The advantage of the first penalty, $J_1$, is that it still requires only one tuning parameter $\lambda$; the disadvantage is the ad hoc use of the regression coefficient $\hat{\phi}^*_{j,j-1}$, which may not be close to $\hat{\phi}_{j,j-1}$, but we can reasonably hope is on the same scale. The second penalty, $J_2$, does not require this extra regression coefficient, but it does require selection of two tuning parameters. It turns out, however, that, in practice, the value of $\lambda_2$ is not as important as that of $\lambda_1$, as the ratio term will be infinite whenever a coefficient in the denominator is shrunk to 0. In practice, on both simulations and real data, we have not found much difference between the three versions $J_0$, $J_1$, and $J_2$, although in general $J_1$ tends to be better than $J_0$, and $J_2$ better than $J_1$. In what follows, we will write $J$ for the three nested penalties $J_0$, $J_1$ and $J_2$ if any one of them can be substituted.

**Adaptive banding for covariance estimation:**

1. For $j = 1$, $\hat{\sigma}_1^2 = \mathrm{Var}(X_1)$.
2. For each $j = 2, \ldots, p$, let

$$(12) \qquad (\hat{\sigma}_j, \hat{\phi}_j) = \underset{\sigma_j, \phi_j}{\arg\min}\, \ell_j(\sigma_j, \phi_j, \mathbf{x}_1, \ldots, \mathbf{x}_n) + J(\phi_j).$$

3. Compute $\hat{\Sigma}^{-1}$ according to (3); let $\hat{\Sigma} = (\hat{\Sigma}^{-1})^{-1}$.

2.2. *The algorithm.* The minimization of (12) is a nontrivial problem, since the penalties $J$ are not convex. We developed an iterative procedure for this minimization, which we found to work well and converge quickly in practice.

The algorithm requires an initial estimate of the coefficients $\phi_j$. In the case $p < n$, one could initialize with coefficients $\hat{\phi}_j$ fitted without a penalty, which are given by the usual least squares estimates from regressing $X_j$ on $X_{j-1}, \ldots, X_1$. If $p > n$, however, these are not defined. Instead, we initialize with $\hat{\phi}_{jt}^{(0)} = \hat{\phi}_{jt}^*$, which are found by regressing $X_j$ on $X_t$ alone, for $t = 1, \ldots, j - 1$. Then we iterate between steps 1 and 2 until convergence.

**Step 1.** Given $\hat{\phi}_j^{(k)}$, solve for $\hat{\sigma}_j^{(k)}$ (the residual sum of squares is given in closed form):

$$(\hat{\sigma}_j^{(k)})^2 = \frac{1}{n} \sum_{i=1}^{n} \left( x_{ij} - \sum_{t=1}^{j-1} \hat{\phi}_{jt}^{(k)} x_{it} \right)^2.$$

**Step 2.** Given $\hat{\phi}_j^{(k)}$ and $\hat{\sigma}_j^{(k)}$, solve for $\hat{\phi}_j^{(k+1)}$. Here we use the following standard local quadratic approximation [also used by [Fan and Li (2001)](#) and [Huang et al. (2006)](#), among others]:

$$(13) \qquad |\phi_{jt}^{(k+1)}| \approx \frac{(\phi_{jt}^{(k+1)})^2}{2|\phi_{jt}^{(k)}|} + \frac{|\phi_{jt}^{(k)}|}{2}.$$

This approximation, together with substituting the previous values $\phi_{jt}^{(k)}$ in the denominator of the ratios in the penalty, converts the minimization into a ridge (quadratic) problem, which can be solved in closed form. For example, for the $J_2$ penalty, we solve

$$\hat{\phi}_j^{(k+1)} = \arg\min_{\phi_j} \frac{1}{(\hat{\sigma}_j^{(k)})^2} \sum_{i=1}^{n} \left( x_{ij} - \sum_{t=1}^{j-1} \phi_{jt} x_{it} \right)^2$$

$$+ \lambda_1 \sum_{t=1}^{j-1} \frac{\phi_{jt}^2}{2|\hat{\phi}_{jt}^{(k)}|} + \lambda_2 \sum_{t=1}^{j-2} \frac{\phi_{jt}^2}{2|\hat{\phi}_{jt}^{(k)}| \cdot |\hat{\phi}_{j,t+1}^{(k)}|}.$$

For numerical stability, we threshold the absolute value of every estimate at $10^{-10}$ over different iterations, and at the end of the iterations, set all estimates equal to $10^{-10}$ to zero. The approximation for $J_0$ and $J_1$ penalties is analogous. The function we are minimizing in (12) is not convex, therefore, there is no guarantee of finding the global minimum. However, in the simulations we have tried, where we know the underlying truth (see Section 3.1 for details), we have not encountered any problems with spurious local minima with the choice of starting values described above.

Another approach in Step 2 above is to use the "shooting" strategy [Fu (1998); Friedman et al. (2007)]. That is, we sequentially solve for $\phi_{jt}$: for each $t = 1, \ldots, j-1$, we fix $\sigma_j$ and $\phi_{-jt} = (\phi_{j1}, \ldots, \phi_{j,t-1}, \phi_{j,t+1}, \ldots, \phi_{j,j-1})^\top$ at their most recent estimates and minimize (12) over $\phi_{jt}$, and iterate until convergence. Since each minimization over $\phi_{jt}$ involves only one parameter and the objective function is piecewise convex, the computation is trivial. Also, since at each iteration the value of the objective function decreases, convergence is guaranteed. In our experience, these two approaches, the local quadratic approximation and the shooting strategy, do not differ much in terms of the computational cost and the solutions they offer.

The algorithm also requires selecting the tuning parameter $\lambda$, or, in the case of $J_2$, two tuning parameters $\lambda_1$ and $\lambda_2$. We selected tuning parameters on a validation set which we set aside from the original training data; alternatively, 5-fold cross-validation can be used. As discussed above, we found that the value of $\lambda_2$ in $J_2$ is not as important; however, in all examples in this paper the computational burden was small enough to optimize over both parameters.

**3. Numerical results.** In this section we compare adaptive banding to other methods of regularizing the inverse. Our primary comparison is with the Lasso method of Huang et al. (2006) and with nonadaptive banding of Bickel and Levina (2007); these methods are closest to ours and also provide a sparse estimate of the Cholesky factor. As a benchmark, we also include the shrinkage estimator of Ledoit and Wolf (2003), which does not depend on the order of variables.

3.1. *Simulation data.* Simulations were carried out for three different covariance models. The first one has a tri-diagonal Cholesky factor and, hence, a tri-diagonal inverse:

$$\boldsymbol{\Sigma}_1 : \phi_{j,j-1} = 0.8; \qquad \phi_{j,j'} = 0, \qquad j' < j - 1; \qquad \sigma_j^2 = 0.01.$$

The second one has entries of the Cholesky factor exponentially decaying as one moves away from the diagonal. Its inverse is not sparse, but instead has many small entries:

$$\boldsymbol{\Sigma}_2 : \phi_{j,j'} = 0.5^{|j-j'|}, \qquad j' < j; \qquad \sigma_j^2 = 0.01.$$

Both these models were considered by Huang et al. (2006), and similar models were also considered by Bickel and Levina (2007). In both $\Sigma_1$ and $\Sigma_2$, all the rows have the same structure, which favors regular nonadaptive banding.

To test the ability of our algorithm to adapt, we also considered the following structure:

$$\boldsymbol{\Sigma_3}: k_j \sim U(1, \lceil j/2 \rceil); \qquad \phi_{j,j'} = 0.5, \qquad k_j \leq j' \leq j-1;$$

$$\phi_{j,j'} = 0, \qquad j' < k_j; \qquad \sigma_j^2 = 0.01.$$

Here $U(k_1, k_2)$ denotes an integer selected at random from all integers from $k_1$ to $k_2$. For moderate values of $p$, this structure is stable, and this is what we generate for $p = 30$ in the simulations below. For larger $p$, some realizations can generate a poorly conditioned true covariance matrix, which is not a problem in principle, but makes computing performance measures awkward. To avoid this problem, we divided the variables for $p = 100$ and $p = 200$ into 3 and 6 independent blocks, respectively, and generated a random structure from the model described above for each of the blocks. We will refer to all these models as $\boldsymbol{\Sigma_3}$. The structure of $\boldsymbol{\Sigma_3}$ should benefit more from adaptive banding.

For each of the covariance models, we generated $n = 100$ training observations, along with a separate set of 100 validation observations. We considered three different values of $p$: 30, 100 and 200, and two different distributions: normal and multivariate $t$ with 3 degrees of freedom, to test the behavior of the estimator on heavy-tailed data. The estimators were computed on the training data, with tuning parameters for all methods selected by maximizing the likelihood on the validation data. Using these values of the tuning parameters, we then computed the estimated covariance matrix on the training data and compared it to the true covariance matrix.

There are many criteria one can use to evaluate covariance matrix estimation, for example, any one of the matrix norms can be calculated for the difference ($L_1$, $L_2$, $L_\infty$, or Frobenius norm). There is no general agreement on which loss to use in which situation. Here we use the Kullback–Leibler loss for the concentration matrix, which was used in Yuan and Lin (2007). The Kullback–Leibler loss is defined as follows:

$$(14) \qquad \Delta_{\mathrm{KL}}(\boldsymbol{\Sigma}, \hat{\boldsymbol{\Sigma}}) = \mathrm{tr}(\hat{\boldsymbol{\Sigma}}^{-1}\boldsymbol{\Sigma}) - \ln|\hat{\boldsymbol{\Sigma}}^{-1}\boldsymbol{\Sigma}| - p.$$

Another popular loss is the entropy loss for the covariance matrix, which was used by Huang et al. (2006). The entropy loss is the same as the Kullback–Leibler loss except the roles of the covariance matrix and its inverse are switched.

The entropy loss can be derived from the Wishart likelihood [Anderson (1958)]. While one does not expect major disagreements between these losses, the entropy loss is a more appropriate measure if the covariance matrix itself is the primary object of interest (as in PCA, e.g.), and the Kullback–Leibler loss is a more direct measure of the estimate of the concentration matrix. Both these losses are not normalized by dimension and therefore cannot be compared directly for different $p$'s. We have also tried matrix norms and the so-called quadratic loss from Huang et al. (2006) and found that, while there is no perfect agreement between results every time, qualitatively they are quite similar. The main conclusions we draw from comparing estimators using the Kullback–Leibler loss would be the same if any other loss had been used.

The results for the normal data and the three models are summarized in Table 1, which gives the average losses and the corresponding standard errors over 50 replications. The NA values for the sample appear when the matrix is singular. The $J_0$ penalty has been omitted because it is dominated by $J_1$ and $J_2$.

In general, we see that banding and adaptive banding perform better on all three models than the sample, Ledoit–Wolf's estimator and Lasso. On $\Sigma_1$ and $\Sigma_2$, as expected, banding and adaptive banding are very similar (particularly once standard errors are taken into account); but on $\Sigma_3$, adaptive banding does better, and the larger $p$, the bigger the difference. Also, for normal data the $J_2$ penalty always dominates $J_1$, though they are quite close.

To test the behavior of the methods with heavy-tailed data, we also performed simulations for the same three covariance models under the multivariate $t_3$ distribution (the heaviest-tail $t$ distribution with finite variance).

TABLE 1

*Multivariate normal simulations for models $\Sigma_1$ (banded Cholesky factor), $\Sigma_2$ (nonsparse Cholesky factor with elements decaying exponentially as one moves away from the diagonal) and $\Sigma_3$ (sparse Cholesky factor with variable length rows). The Kullback–Leibler losses (means and, in parentheses, standard errors for the means over 50 replications) are reported for sample covariance, the shrinkage estimator of Ledoit and Wolf (2003), the Lasso method of Huang et al. (2006), the nonadaptive banding method of Bickel and Levina (2007), and our adaptive banding with penalties $J_1$ and $J_2$*

| $p$ | Sample | Ledoit–Wolf | Lasso | $J_1$ | $J_2$ | Banding |
|-----|--------|-------------|-------|-------|-------|---------|
| | | | $\Sigma_1$ | | | |
| 30 | 8.38(0.14) | 3.59(0.04) | 1.26(0.04) | 0.79(0.02) | 0.64(0.02) | 0.63(0.02) |
| 100 | NA | 29.33(0.12) | 6.91(0.11) | 2.68(0.04) | 2.21(0.03) | 2.21(0.03) |
| 200 | NA | 90.86(0.19) | 14.57(0.13) | 5.10(0.06) | 4.35(0.05) | 4.34(0.05) |
| | | | $\Sigma_2$ | | | |
| 30 | 8.38(0.14) | 3.59(0.02) | 2.81(0.04) | 1.42(0.03) | 1.32(0.02) | 1.29(0.03) |
| 100 | NA | 18.16(0.02) | 16.12(0.09) | 5.01(0.07) | 4.68(0.06) | 4.55(0.05) |
| 200 | NA | 40.34(0.02) | 32.84(0.11) | 9.88(0.06) | 9.28(0.06) | 8.95(0.06) |
| | | | $\Sigma_3$ | | | |
| 30 | 8.68(0.12) | 171.31(1.00) | 4.62(0.07) | 3.26(0.05) | 3.14(0.06) | 3.82(0.05) |
| 100 | NA | 945.65(2.15) | 35.60(0.71) | 11.82(0.13) | 11.24(0.12) | 14.34(0.09) |
| 200 | NA | 1938.32(3.04) | 118.84(1.54) | 23.30(0.17) | 22.70(0.16) | 29.50(0.14) |

These results are given in Table 2. All methods perform worse than they do for normal data, but banding and adaptive banding still do better than other methods. Because the standard errors are larger, it is harder to establish a uniform winner among $J_1$, $J_2$ and banding, but generally these results are consistent with results obtained for normal data.

Finally, we note that the differences between estimators are amplified with growing dimension $p$: while the patterns remain the same for all three values of $p$ considered (30, 100 and 200), quantitatively the improvement of adaptive banding over the Ledoit–Wolf estimator and Lasso is the largest at $p = 200$, and is expected to be even more for higher dimensions.

Since one advantage of adaptive banding as compared to Lasso is preserving sparsity in the inverse itself, we also compared percentages of true zeros both in the Cholesky factor and in the inverse that were estimated as zeros, for the models $\mathbf{\Sigma_1}$ and $\mathbf{\Sigma_3}$ ($\mathbf{\Sigma_2}$ is not sparse). The results are shown in Table 3. While for the easier case of $\mathbf{\Sigma_1}$ all methods do a reasonable job of finding zeros in the Cholesky factor, Lasso loses them in the inverse, whereas both kinds of banding do not. This effect is even more apparent on the more challenging case of $\mathbf{\Sigma_3}$.

To gain additional insight into the sparsity of structures produced by the different methods, we also show heatmap plots of the percentage of times each entry of the Cholesky factor (Figure 2) and the inverse itself (Figure 3) were estimated as zeros. It is clear that only adaptive banding reflects the true underlying structure.

Overall, the simulations show that the adaptive banding achieves the goals that it was designed for: it has more flexibility than banding, and therefore is better able to capture the underlying sparse structure, but, unlike the

TABLE 2
*Multivariate $t_3$ simulations for models $\mathbf{\Sigma_1}$, $\mathbf{\Sigma_2}$, $\mathbf{\Sigma_3}$. Descriptions for the entries are the same as those in Table 1*

| $p$ | Sample | Ledoit–Wolf | Lasso | $J_1$ | $J_2$ | Banding |
|---|---|---|---|---|---|---|
| | | | $\mathbf{\Sigma_1}$ | | | |
| 30 | 30.33(0.65) | 9.22(0.65) | 7.60(0.74) | 4.32(0.21) | 3.68(0.19) | 4.22(0.60) |
| 100 | NA | 58.24(2.61) | 38.99(1.44) | 15.58(0.78) | 13.85(0.72) | 13.74(0.72) |
| 200 | NA | 139.21(3.02) | 111.62(2.73) | 31.45(1.80) | 28.22(1.71) | 27.95(1.70) |
| | | | $\mathbf{\Sigma_2}$ | | | |
| 30 | 30.33(0.65) | 6.20(0.15) | 8.44(0.20) | 5.91(0.24) | 5.21(0.22) | 5.23(0.24) |
| 100 | NA | 24.37(0.67) | 31.92(0.83) | 21.76(0.76) | 18.87(0.71) | 19.33(0.85) |
| 200 | NA | 50.40(1.41) | 64.28(1.98) | 44.58(2.00) | 38.46(1.75) | 39.81(1.98) |
| | | | $\mathbf{\Sigma_3}$ | | | |
| 30 | 30.77(0.74) | 199.73(4.32) | 14.48(0.40) | 11.47(0.44) | 11.57(0.47) | 11.69(0.39) |
| 100 | NA | 1061.54(12.62) | 82.05(1.47) | 43.38(1.14) | 45.01(1.13) | 42.78(1.04) |
| 200 | NA | 2182.54(21.29) | 182.82(9.51) | 87.5(2.75) | 91.25(2.79) | 85.65(2.49) |

TABLE 3
*Percentage of true zeros in the Cholesky factor and in the inverse estimated as zeros for multivariate normal data, for models $\mathbf{\Sigma_1}$ and $\mathbf{\Sigma_3}$ (means and, in parentheses, standard errors for the means over 50 replications)*

| | Zeros in the Cholesky factor (%) | | | | Zeros in $\mathbf{\Sigma^{-1}}$ (%) | | | |
|---|---|---|---|---|---|---|---|---|
| $p$ | Lasso | $J_1$ | $J_2$ | Banding | Lasso | $J_1$ | $J_2$ | Banding |
| | | | | $\mathbf{\Sigma_1}$ | | | | |
| 30 | 70.5(0.4) | 94.5(0.3) | 96.3(0.4) | 100(0) | 31.4(0.8) | 94.5(0.3) | 96.3(0.4) | 100(0) |
| 100 | 92.7(0.1) | 98.6(0.04) | 99.1(0.1) | 100(0) | 76.4(0.5) | 98.6(0.3) | 99.1(0.04) | 100(0) |
| 200 | 93.7(0.06) | 99.3(0.01) | 99.5(0.04) | 100(0) | 69.9(0.4) | 99.3(0.01) | 99.5(0.04) | 100(0) |
| | | | | $\mathbf{\Sigma_3}$ | | | | |
| 30 | 55.6(1.5) | 83.2(0.5) | 80.9(0.7) | 72.1(0.9) | 10.2(0.7) | 75.3(0.9) | 70.4(1.5) | 73.1(0.7) |
| 100 | 88.3(0.1) | 94.9(0.1) | 94.9(0.1) | 92.8(0.2) | 55.1(0.5) | 92.3(0.3) | 92.3(0.2) | 93.5(0.2) |
| 200 | 92.4(0.1) | 97.6(0.04) | 97.7(0.04) | 96.7(0.1) | 84.4(0.9) | 96.6(0.1) | 96.7(0.1) | 97.1(0.1) |

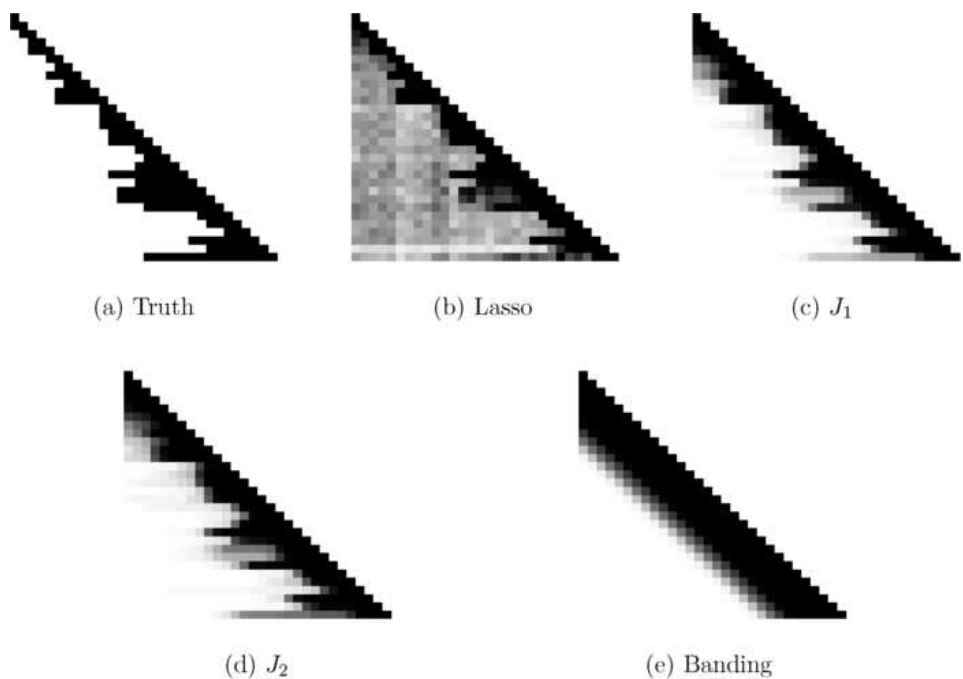

(a) Truth        (b) Lasso        (c) $J_1$

(d) $J_2$        (e) Banding

FIG. 2. *Heatmap plots of percentage of zeros at each location in the inverse (out of 50 replications) for $\mathbf{\Sigma_3}, p = 30$. Black represents 100%, white 0%.*

Lasso, it has the ability to preserve the sparsity in the inverse as well as in the Cholesky factor.

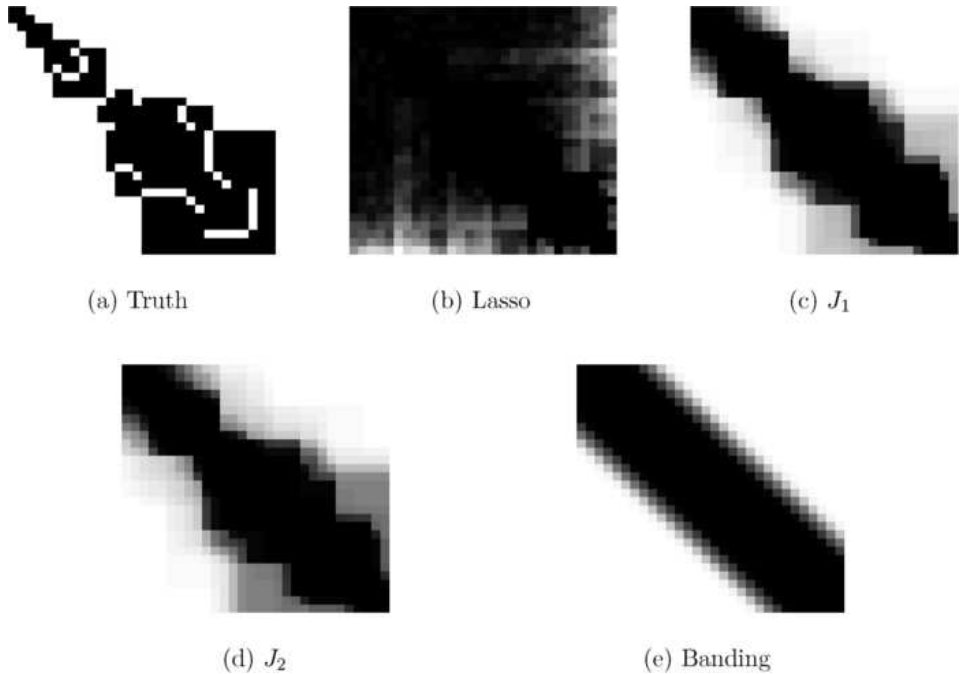

Fig. 3. *Heatmap plots of percentage of zeros at each location in the inverse (out of 50 replications) for* $\mathbf{\Sigma_3}$, $p = 30$. *Black represents 100%, white 0%.*

3.2. *Prostate cancer data.* In this section we consider an application to a prostate cancer dataset [Adam et al. (2002)]. The current standard screening approach for prostate cancer is a serum test for a prostate-specific antigen. However, the accuracy is far from satisfactory [Pannek and Partin (1998) and Djavan et al. (1999)], and it is believed that a combination or a panel of biomarkers will be required to improve the detection of prostate cancer [Stamey et al. (2002)]. Recent advances in high-throughput mass spectroscopy have allowed one to simultaneously resolve and analyze thousands of proteins. In protein mass spectroscopy, we observe, for each blood serum sample $i$, the intensity $x_{ij}$ for many time-of-flight values. Time of flight is related to the mass over charge ratio $m/z$ of the constituent proteins in the blood. The full dataset we consider [Adam et al. (2002)] consists of 157 healthy patients and 167 with prostate cancer. The goal is to discriminate between the two groups. Following the original researchers, we ignored $m/z$-sites below 2000, where chemical artifacts can occur. To smooth the intensity profile, we average the data in consecutive blocks of 10, giving a total of 218 sites. Thus, each observation $\mathbf{x} = (x_1, \ldots, x_{218})$ consists of an intensity profile of length 218, with a known class (cancer or noncancer) membership. The prostate cancer dataset we consider comes with pre-specified training ($n = 216$) and test sets ($N = 108$). Figure 4 displays the mean intensities for "cancer" and "noncancer" from the training data.

We consider the linear discriminant method (LDA) and the quadratic discriminant method (QDA). The linear and quadratic discriminant analysis assume the class-conditional density of $\mathbf{x}$ in class $k$ is normal $\mathcal{N}(\mu_k, \Sigma_k)$. The LDA arises in the special case when one assumes that the classes have a common covariance matrix $\Sigma_k = \Sigma, \forall k$. If the $\Sigma_k$ are not assumed to be equal, we then get QDA. The linear and quadratic discriminant scores are as follows:

$$\text{LDA}: \delta_k(\mathbf{x}) = \mathbf{x}^\top \hat{\Sigma}^{-1} \hat{\mu}_k - \tfrac{1}{2} \hat{\mu}_k^\top \hat{\Sigma}^{-1} \hat{\mu}_k + \log \hat{\pi}_k,$$

$$\text{QDA}: \delta_k(\mathbf{x}) = -\tfrac{1}{2} \log |\hat{\Sigma}_k| - \tfrac{1}{2}(\mathbf{x} - \hat{\mu}_k)^\top \hat{\Sigma}_k^{-1}(\mathbf{x} - \hat{\mu}_k) + \log \hat{\pi}_k,$$

where $\hat{\pi}_k = n_k/n$ is the proportion of the number of class-$k$ observations in the training data, and the classification rule is given by $\arg\max_k \delta_k(\mathbf{x})$. Detailed information on LDA and QDA can be found in Mardia et al. (1979).

Using the training data, we estimate

$$\hat{\mu}_k = \frac{1}{n_k} \sum_{i \in \text{class } k} \mathbf{x}_i$$

and estimate $\hat{\Sigma}^{-1}$ or $\hat{\Sigma}_k^{-1}$ using five different methods: the shrinkage toward the identity estimator of Ledoit and Wolf (2003), banding the Cholesky factor, the Lasso estimator, and our adaptive banding method; we also include the Naive Bayes method as a benchmark, since it corresponds to LDA with the covariance matrix estimated by a diagonal matrix. Tuning parameters in different methods are chosen via five-fold cross-validation based on the training data. Mean vectors and covariance matrices were estimated on the training data and plugged into the classification rule, which was then applied to the test data. Note that for this dataset $p$ is greater than $n$, hence,

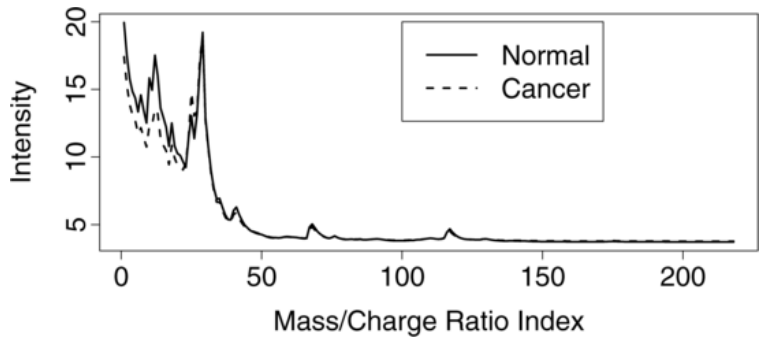

FIG. 4. *The mean intensity for "cancer" and "noncancer" from the training data.*

the sample covariance matrix is not invertible and cannot be used in LDA or QDA.

The results as measured by the test set classification error are summarized in Table 4. For this particular dataset, we can see that overall the QDA performs much worse than the LDA, so we focus on the results of the LDA. The Naive Bayes method, which assumes independence among variables, has the worst performance. Banding (with a common bandwidth) and the Lasso method perform similarly and better than the Naive Bayes method. Our adaptive banding method performs the best, with either the $J_1$ or the $J_2$ penalty. To gain more insight about the sparsity structures of different estimators, we plot the structures of the estimated Cholesky factors and the corresponding $\hat{\Sigma}^{-1}$ of different methods in Figures 5 and 6 (black represents nonzero, and white represents zero). Based on the differences in the classification performance, these plots imply that the Lasso method may have included many unimportant elements in $\hat{\Sigma}^{-1}$ (estimating zeros as nonzeros), while the banding method may have missed some important elements (estimating nonzeros as zeros). The estimated Cholesky factors and the corresponding $\hat{\Sigma}^{-1}$'s from the adaptive banding method ($J_1$ and $J_2$) represent an interesting structure: the intensities at higher $m/z$-values are more or less conditionally independent, while the intensities at lower $m/z$-values show a "block-diagonal" structure.

We note that in the above analysis we used likelihood on the cross-validation data as the criterion for selecting tuning parameters. As an alternative, we also considered using the classification error as the selection criterion. The results from the two selection criteria are similar. For simplicity of exposition, we only presented results from using the likelihood as the selection criterion.

Finally, we note that Adam et al. (2002) reported an error rate around 5% for a four-class version of this problem, using a peak finding procedure followed by a decision tree algorithm. However, we have had difficulty replicating their results, even when using their extracted peaks. In Tibshirani et al. (2005) the following classification errors were reported for other methods applied to the two-class dataset we used here: 30 for Nearest Shrunken Centroids [Tibshirani et al. (2003)], and 16 for both Lasso and fused Lasso [Tibshirani et al. (2005)]. However, note that these authors did not apply block smoothing.

TABLE 4
*Test errors (out of 108 samples) on the prostate cancer dataset*

| Method | Naive Bayes | Ledoit & Wolf | Banding | Lasso | $J_1$ | $J_2$ |
|---|---|---|---|---|---|---|
| Test error (LDA) | 32 | 16 | 16 | 18 | 11 | 12 |
| Test error (QDA) | 32 | 51 | 46 | 49 | 31 | 29 |

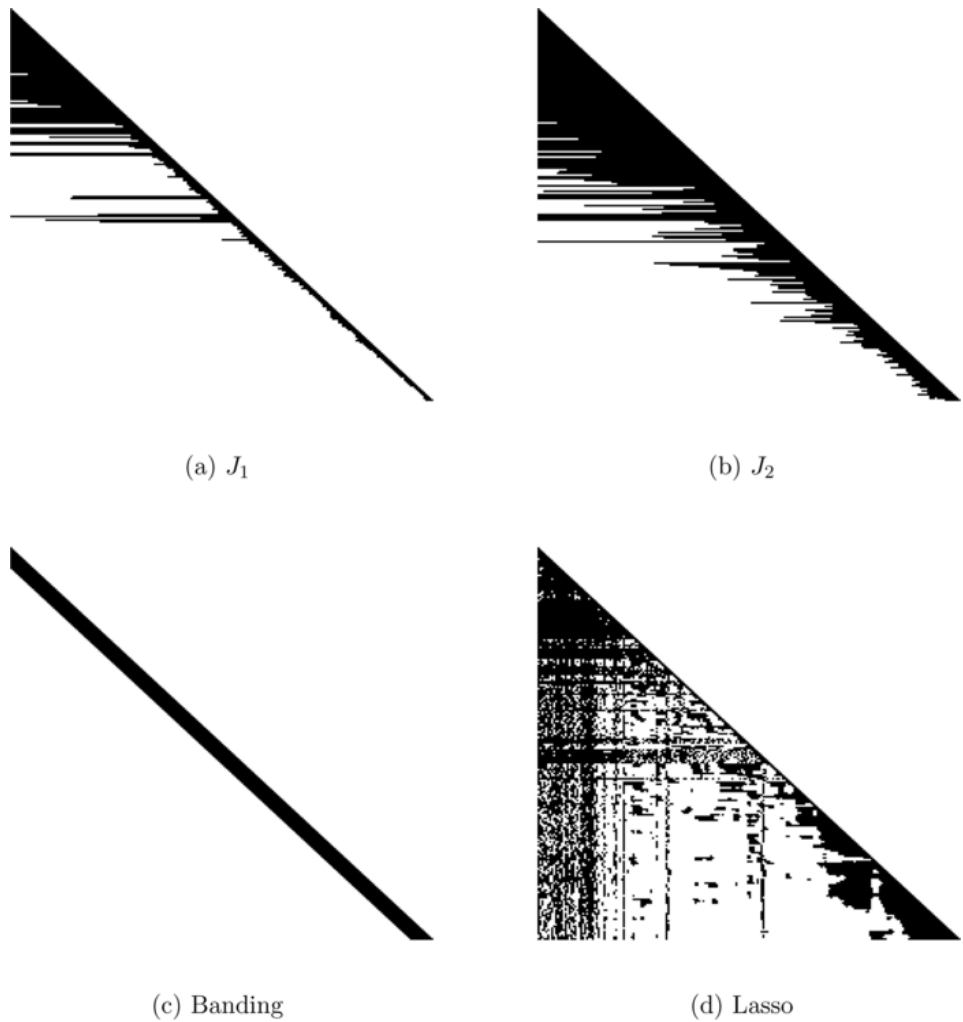

(a) $J_1$                         (b) $J_2$

(c) Banding                    (d) Lasso

FIG. 5. *Structure of Cholesky factors estimated for the prostate data. White corresponds to zero, black to nonzero.*

**4. Summary and discussion.** We have presented a new covariance estimator for ordered variables with a banded structure, which, by selecting the bandwidth adaptively for each row of the Cholesky factor, achieves more flexibility than regular banding but still preserves sparsity in the inverse. Adaptive banding is achieved using a novel nested Lasso penalty, which takes into account the ordering structure among the variables. The estimator has been shown to do well both in simulations and a real data example. Zhao et al. (2006) proposed a related penalty, the composite absolute penalty (CAP), for handling hierarchical structures in variables. However,

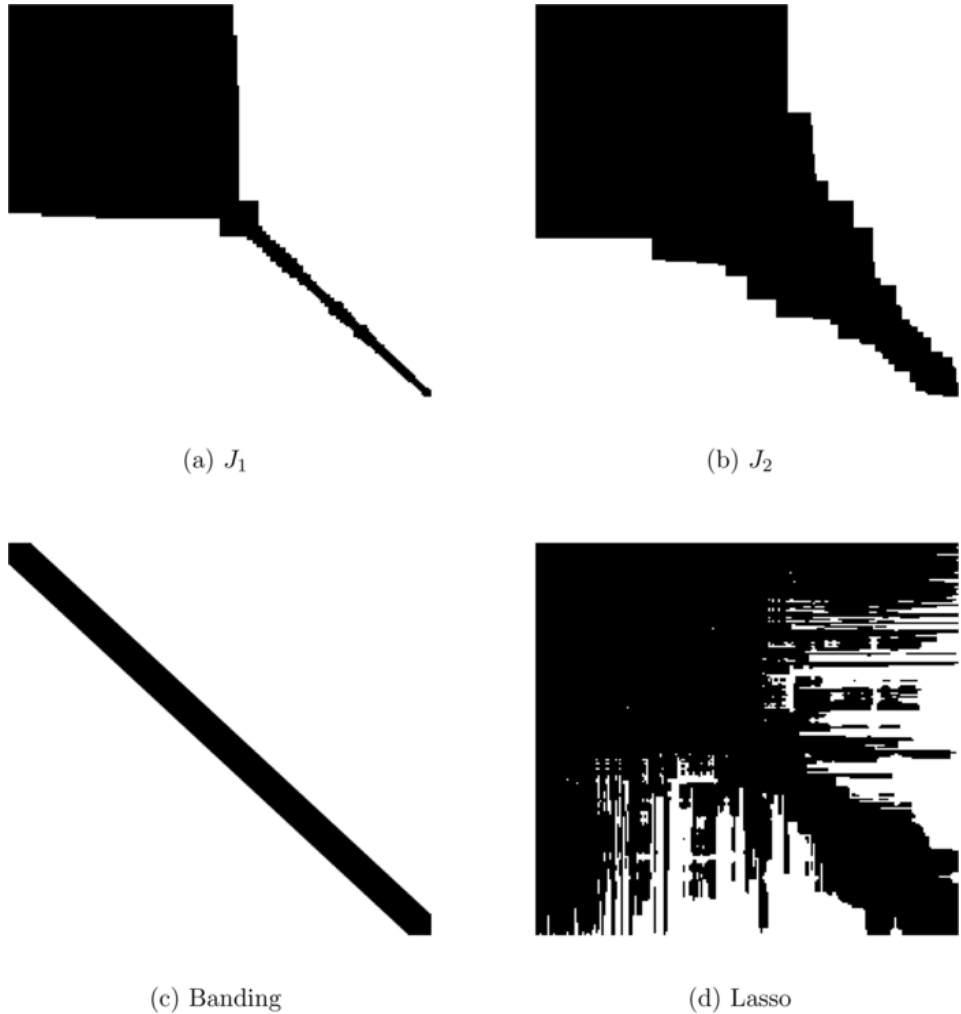

(a) $J_1$                (b) $J_2$

(c) Banding          (d) Lasso

Fig. 6. *Structure of the inverse covariance matrix estimated for the prostate data. White corresponds to zero, black to nonzero.*

Zhao et al. (2006) only considered a hierarchy with two levels, while, in our setting, there are essentially $p - 1$ hierarchical levels; hence, it is not clear how to directly apply CAP without dramatically increasing the number of tuning parameters.

The theoretical properties of the estimator are a subject for future work. The nested Lasso penalty is not convex in the parameters; it is likely that the theory developed by Fan and Li (2001) for nonconvex penalized maximum likelihood estimation can be extended to cover the nested Lasso (it is not directly applicable since our penalty cannot be decomposed into a sum of identical penalties on the individual coefficients). However, that theory was developed only for the case of fixed $p$, $n \to \infty$, and the more relevant analysis for estimation of large covariance matrices would be under the assumption $p \to \infty$, $n \to \infty$, with $p$ growing at a rate equal to or possibly faster than that of $n$, as was done for the banded estimator by Bickel and Levina (2007). Another interesting question for future work is extending this idea to estimators invariable under variable permutations.

**Acknowledgments.** We thank the Editor, Michael Stein, and two referees for their feedback which led us to improve the paper, particularly the data section. We also thank Jianqing Fan, Xiaoli Meng and Bin Yu for helpful comments.

E. Levina
A. Rothman
J. Zhu
Department of Statistics
University of Michigan
Ann Arbor, Michigan 48109–1107
USA
E-mail: elevina@umich.edu
        ajrothma@umich.edu
        jizhu@umich.edu